\documentclass[aps,prl,reprint,superscriptaddress,longbibliography]{revtex4-2}
\usepackage[T1]{fontenc}
\usepackage{amsmath,amssymb,amsthm}
\usepackage{mathtools,booktabs,tikz,microtype}
\usepackage{xcolor}
\definecolor{apsblue}{rgb}{0.18, 0.19, 0.57}
\usepackage[colorlinks=true,allcolors=apsblue, breaklinks=true]{hyperref}
\usepackage{graphicx}
\usepackage{subcaption}

\makeatletter
\DeclareFontFamily{OMX}{MnSymbolE}{}
\DeclareSymbolFont{MnLargeSymbols}{OMX}{MnSymbolE}{m}{n}
\SetSymbolFont{MnLargeSymbols}{bold}{OMX}{MnSymbolE}{b}{n}
\DeclareFontShape{OMX}{MnSymbolE}{m}{n}{
    <-6>  MnSymbolE5
   <6-7>  MnSymbolE6
   <7-8>  MnSymbolE7
   <8-9>  MnSymbolE8
   <9-10> MnSymbolE9
  <10-12> MnSymbolE10
  <12->   MnSymbolE12
}{}
\DeclareFontShape{OMX}{MnSymbolE}{b}{n}{
    <-6>  MnSymbolE-Bold5
   <6-7>  MnSymbolE-Bold6
   <7-8>  MnSymbolE-Bold7
   <8-9>  MnSymbolE-Bold8
   <9-10> MnSymbolE-Bold9
  <10-12> MnSymbolE-Bold10
  <12->   MnSymbolE-Bold12
}{}

\let\llangle\@undefined
\let\rrangle\@undefined
\DeclareMathDelimiter{\llangle}{\mathopen}%
                     {MnLargeSymbols}{'164}{MnLargeSymbols}{'164}
\DeclareMathDelimiter{\rrangle}{\mathclose}%
                     {MnLargeSymbols}{'171}{MnLargeSymbols}{'171}
\makeatother

\begin{document}
\title{Giant Graviton Correlators:\\ From Non-maximal to Dual, Bootstrap and Hidden Symmetry}
\author{Qi Chen}
\affiliation{Kavli Institute for Theoretical Sciences, University of
Chinese Academy of Sciences, Beijing 100190, China}
\author{Zhongjie Huang}
\affiliation{Kavli Institute for Theoretical Sciences, University of
Chinese Academy of Sciences, Beijing 100190, China}
\author{Yunfeng Jiang}
\affiliation{School of physics \& Shing-Tung Yau Center, Southeast University, Nanjing, China}
\author{Xinan Zhou}
\affiliation{Kavli Institute for Theoretical Sciences, University of
Chinese Academy of Sciences, Beijing 100190, China}
\date[]{} 
\begin{abstract}
    We compute four-point functions of two light $\frac{1}{2}$-BPS operators and two heavy giant graviton operators in $\mathcal{N}=4$ super Yang--Mills theory at large $N$ and infinite 't Hooft coupling. Using bootstrap techniques, we determine the correlators for arbitrary light-operator dimensions and arbitrary allowed sizes of non-maximal sphere and dual giant gravitons. We show that these correlators exhibit a hidden ten-dimensional conformal symmetry, allowing them to be organized into compact generating functions. Remarkably, the generating functions for both types of giant gravitons can be obtained from the maximal giant case through $SO(2,10)$ rotations in the embedding space.
\end{abstract}

\maketitle

\section{Introduction}
Understanding nonplanar physics remains a major challenge in $\mathcal{N}=4$ super Yang--Mills (SYM) theory. While the planar limit has led to remarkable progress, much less is known beyond planarity, especially at strong coupling. Giant gravitons \cite{McGreevy:2000cw,Balasubramanian:2001nh,Hashimoto:2000zp} provide particularly useful probes of this regime. They are heavy $\frac{1}{2}$-BPS operators and are dual to D3-branes in AdS$_5\times$S$^5$. Their conformal dimensions scale with $N$, making the conventional planar expansion inadequate even in the large-$N$ limit. Recently, significant progress was made in computing four-point functions involving two maximal giant gravitons and two light $\frac{1}{2}$-BPS operators. By treating the heavy operators as a defect system and extending the bootstrap techniques of \cite{Rastelli:2016nze,Rastelli:2017udc}, these correlators were systematically determined at large $N$ and infinite 't Hooft coupling \cite{Chen:2025yxg,Chen:2026ium}. Remarkably, the correlators exhibit a hidden ten-dimensional conformal symmetry, generalizing analogous structures previously discovered in defect-free cases \cite{Caron-Huot:2018kta,Rastelli:2019gtj,Alday:2021odx}. 

A natural question is whether these results can be extended to more general giant gravitons. There are two distinct families: non-maximal sphere giants, which wrap an S$^3$ inside S$^5$ and have bounded sizes, and dual giants, which wrap an S$^3$ inside AdS$_5$ and have unbounded sizes. Moving away from the maximal case, however, introduces significant new challenges. First, rather than bootstrapping correlators at a special fixed size, we now  seek entire families of correlators which nontrivially depend on the continuous size parameter. Furthermore, both non-maximal and dual giants possess an additional orbital collective coordinate which generates a richer effective defect spectrum. For dual giants, the situation is even more complicated: their worldvolume extends into AdS$_5$, giving rise to new classes of Witten diagrams which are substantially more involved than those for sphere giants. 

In this Letter, we overcome these difficulties and determine all four-point correlators of two light $\frac{1}{2}$-BPS operators and two non-maximal sphere or dual giant gravitons, for arbitrary light-operator dimensions and allowed giant graviton sizes in the supergravity limit. We achieve this by extending the defect bootstrap method to incorporate  orbital averaging and new classes of Witten diagrams. The correlators turn out to exhibit remarkable structures. Despite the substantial differences between the two types of giant gravitons, their correlators both exhibit a more elaborate version of the hidden ten-dimensional conformal structure, allowing them to be compactly organized into generating functions. A crucial new ingredient in the uplifting prescription is the averaging over the orbital collective coordinate. Even more surprisingly, the generating functions of both non-maximal and dual giants can be obtained from the maximal case through $SO(2,10)$ rotations in the total embedding space. Consequently, the dependence on both the light-operator dimensions and the types and sizes of giant gravitons is governed by a unified geometric origin, revealing an unexpected universality.

\section{Superconformal kinematics}
We consider four-point functions of $\frac{1}{2}$-BPS operators $\langle \mathcal{O}_{k_1} \mathcal{O}_{k_2} \mathcal{G}_M \mathcal{G}_M \rangle$.
The light single-particle supergravitons
\begin{equation}
    \mathcal{O}_k=\mathcal{N}_{\mathcal{O}_k} {\rm tr}\left(\Phi\cdot t\right)^k+\ldots\;,
\end{equation}
 have dimensions $k=2,3,\ldots$, and are unit-normalized by $\mathcal{N}_{\mathcal{O}_k}$, with $\ldots$ being the multi-trace completions. The six-dimensional null vector $t$ with $t\cdot t=0$ ensures $\mathcal{O}_k$ transforms in the rank-$k$ symmetric traceless representation of $SO(6)_R$. The identical heavy operators are either sphere giant gravitons or dual giant gravitons
\begin{equation}
    \mathcal{G}_M=\frac{1}{M!}\sum_{\sigma\in S_M} \left\{\begin{array}{c} {\rm sgn}(\sigma) {\rm tr}_\sigma (\Phi\cdot t)\;, \quad \text{sphere}\,,\\  {\rm tr}_\sigma (\Phi\cdot t)\;,\quad\quad\quad \text{dual}\,. \end{array}\right.
\end{equation}
They are $\frac{1}{2}$-BPS with dimension $M$ of order $N$ and transform in the rank-$M$ symmetric traceless representation. $M$ is bounded by $M\leq N$ for sphere giants with $M=N$ being  maximal, and is unbounded for dual giants. Holographically, both types of giant gravitons are dual to a D3-brane wrapping an S$^3$. The wrapped S$^3$ is inside S$^5$ for sphere giants and inside AdS$_5$ for dual giants.

In \cite{Chen:2025yxg,Chen:2026ium}, it was pointed out that these correlators are most naturally viewed as defect two-point functions
\begin{equation}
    \llangle \mathcal{O}_{k_1} \mathcal{O}_{k_2} \rrangle=\frac{\langle \mathcal{O}_{k_1} \mathcal{O}_{k_2} \mathcal{G}_M \mathcal{G}_M \rangle}{\langle \mathcal{G}_M \mathcal{G}_M \rangle}\equiv F_{k_1k_2}\;,
\end{equation}
where the heavy operator pair creates a zero-dimensional defect. The defect breaks half of the superconformal symmetry. Particularly, the $SO(2,4)$ conformal symmetry is broken to $SO(3,1)\times SO(1,1)$, and the $SO(6)$ R-symmetry is broken to $SO(4)\times U(1)$. From this defect perspective, it is useful to define projectors. For the spacetime part, we have
\begin{equation}
\mathbb{N}=\frac{P_3\otimes P_4+P_4\otimes P_3}{P_3\cdot P_4}\;,
\end{equation}
where $P=(\frac{1+x^2}{2},\frac{1-x^2}{2},x^\mu)$ are the embedding space vectors. The R-symmetry projector is defined as 
\begin{equation}\label{defM}
\mathbb{M}={\bf 1}_R-\frac{t_3\otimes t_4+t_4\otimes t_3}{t_3\cdot t_4}\;,
\end{equation}
where ${\bf 1}_R$ is the 6d internal space identity. After extracting a factor of one-point functions
\begin{equation}
    F_{k_1k_2}=\prod_{i=1,2}\frac{(\frac{1}{2}t_i\cdot \mathbb{M}\cdot t_i)^{\frac{k_i}{2}} }{(P_i\cdot \mathbb{N}\cdot P_i)^{\frac{k_i}{2}}}\mathcal{F}_{k_1k_2}(z,\bar{z},\alpha,\bar{\alpha})\;,
\end{equation}
the defect correlators depend only on the cross ratios
\begin{equation}
\begin{split}
  {}&  U=\frac{x_{12}^2x_{34}^2}{x_{13}^2x_{24}^2}=z\bar{z}\;,\quad V=\frac{x_{14}^2x_{23}^2}{x_{13}^2x_{24}^2}=(1-z)(1-\bar{z})\;,\\
  {}& \sigma=\frac{t_{13}t_{24}}{t_{12}t_{34}}=\alpha\bar{\alpha}\;,\quad \tau=\frac{t_{14}t_{23}}{t_{12}t_{34}}=(1-\alpha)(1-\bar{\alpha})\;,
\end{split}
\end{equation}
with $x_{ij}=x_i-x_j$ and $t_{ij}=t_i\cdot t_j$. They are further constrained by the superconformal Ward identities \cite{Eden:2000bk,Nirschl:2004pa}
\begin{equation}\label{scfWardid}
    \partial_{\bar{z}} \mathcal{F}_{k_1k_2}(z,\bar{z},\alpha,1/\bar{z})=0\;,
\end{equation}
which can be solved as
\begin{equation}\label{solscfWardid}
    F_{k_1k_2}=F_{k_1k_2}^{\rm free}+\frac{t_{12}^2 x_{13}^4x_{24}^4}{x_{34}^4} R\, H_{k_1k_2}\;,
\end{equation}
with $R=(1-z\alpha)(1-z\bar{\alpha})(1-\bar{z}\alpha)(1-\bar{z}\bar{\alpha})$. Here $F_{k_1k_2}^{\rm free}$ is the free theory defect correlator; the reduced correlator $H_{k_1k_2}$ has shifted R-symmetry charges $k_i-2$ and dimensions $k_i+2$. In particular, $H_{22}$ is an R-symmetry singlet.

\section{Giant graviton geometries}
Before performing the bootstrap calculations, let us first recall how giant gravitons are embedded in the bulk geometry. We  divide the total embedding space $\mathbb{R}^{2,10}$ into $\mathbb{R}^{2,4}\oplus\mathbb{R}^6$ which are correspondingly the embedding space of AdS$_5$ and S$^5$. We single out two orthonormal time-like directions $A_0$, $A_1$ in the AdS factor
\begin{equation}
    A_0^2=A_1^2=-1\;,\quad A_0\cdot A_1=0\;,
\end{equation}
and further choose four spacelike orthonormal directions $G_1,\ldots, G_4$, with $A_i\cdot G_j=0$. Then we have $\mathbb{R}^{2,4}={\rm span}\{A_0,A_1,G_1,G_2,G_3,G_4\}$. Similarly, we select two distinguished internal directions 
\begin{equation}
    B_0^2=B_1^2=1\;,\quad B_0\cdot B_1=0\;,
\end{equation}
which span the giant graviton's orbital plane. The remaining four directions $F_1,\ldots,F_4$ are orthonormal among themselves and orthogonal to $B_0$, $B_1$. We similarly have $\mathbb{R}^6={\rm span}\{B_0,B_1,F_1,F_2,F_3,F_4\}$. We also introduce
\begin{equation}
\begin{split}
B_0(\phi)={}&\cos\phi B_0+\sin\phi B_1\;,\\
B_1(\phi)={}&-\sin\phi B_0+\cos\phi B_1\;.
\end{split}
\end{equation}

\begin{figure}[h]
    \centering
\includegraphics[width=0.28\textwidth]{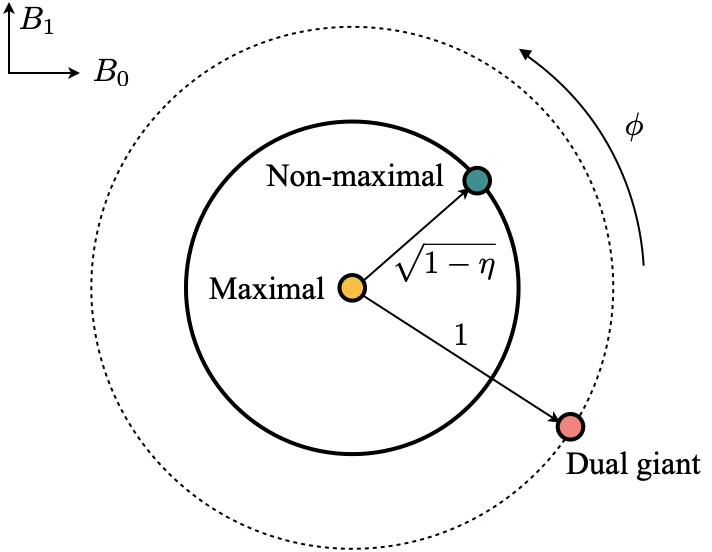}
    \caption{Projection of giant gravitons on the orbital plane $(B_0,B_1)$. As $\phi$ is changed, non-maximal and dual giant gravitons trace out respectively a circle of radius $\sqrt{1-\eta}$ and $1$. The maximal case sits at the center of the plane.}
    \label{fig:orbitalplane}
\end{figure}

For non-maximal giants, we consider the 12d vectors
\begin{equation}
E_0^{\rm S}=(A_0,\sqrt{1-\eta}B_0(\phi))\,,\; E_1^{\rm S}=(A_1,\sqrt{1-\eta}B_1(\phi))\;,
\end{equation}
which satisfy 
\begin{equation}
E_i^{\rm S}\cdot E_j^{\rm S}=-\eta\, \delta_{ij}\;,
\end{equation}
with $\eta=M_{\rm S}/N$. We also regard $F_i$ as 12d vectors via $(0,F_i)$. Together, they span a 6d plane where the giant graviton lives 
\begin{equation}
\mathcal{S}^{\rm S}_{\eta,\phi}={\rm span}\{E_0^{\rm S},E_1^{\rm S},F_1,F_2,F_3,F_4\}\;.
\end{equation}
A point on the giant graviton brane can be written as 
\begin{equation}
\nonumber Z_{\rm S}(t,n)=\cos t E_0^{\rm S}+\sin t E_1^{\rm S}+\sqrt{\eta}\sum_{i=1}
^4 n_i F_i\;,\;\; \sum_i n_i^2=1\;,
\end{equation}
where spacetime and internal components are
\begin{eqnarray}
 && \!\!\!\!\!\!\!\! X(t)=\cos t A_0+\sin t A_1\;, \\
\nonumber && \!\!\!\!\!\! \!\!Y(t,\! n)\!=\!\!\sqrt{1\!-\!\eta}\left(\cos(t\!+\!\phi)B_0\!+\!\sin(t\!+\!\phi) B_1\right)\!+\!\sqrt{\eta}\sum_i \!n_i F_i\,,
\end{eqnarray}
with $X^2=-Y^2=-1$. Here it is clear that $\phi$ is a $U(1)$ modulus of the classical solution describing how the giant graviton D3-brane is embedded. For the special maximal case with $\eta=1$, there is no $\phi$ dependence, see Fig. \ref{fig:orbitalplane}. We can also check that the induced metric is
\begin{equation}
    ds^2_{\rm S}=\eta(-dt^2+d\Omega_3^2)\;.
\end{equation}

 Similarly, for dual giant gravitons we define 
\begin{equation}
E_0^{\rm D}=(\sqrt{1+\beta}A_0,B_0(\phi))\;,\; E_1^{\rm D}=(\sqrt{1+\beta}A_1,B_1(\phi))\;,
\end{equation}
which have the inner product
\begin{equation}
E_i^{\rm D}\cdot E_j^{\rm D}=-\beta\delta_{ij}\;.
\end{equation}
Here $\beta=M_{\rm D}/N$. Instead of $F_i$, we should take the vectors $G_i$ and the dual giant graviton lies in the 6d plane
\begin{equation}
\mathcal{S}_{\beta,\phi}^{\rm D}={\rm span}\{E_0^{\rm D},E_1^{\rm D},G_1,G_2,G_3,G_4\}\;.
\end{equation}
A point on the dual giant graviton is parameterized by 
\begin{equation}
\nonumber Z_{\rm D}(t,n)=\cos t E_0^{\rm D}+\sin t E_1^{\rm D}+\sqrt{\beta}\sum_{i=1}
^4 n_i G_i\;,
\end{equation}
where we have the same orbital phase angle $\phi$ via $B_i(\phi)$, see Fig. \ref{fig:orbitalplane}. The induced metric is
\begin{equation}
ds_{\rm D}^2=\beta(-dt^2+d\Omega_3^2)\;.
\end{equation}

\section{Bootstrap method}

We now use the bootstrap method of \cite{Chen:2025yxg,Chen:2026ium} to compute the correlators. As in the maximal case, we construct an ansatz as a linear combination of contact and exchange Witten diagrams in the bulk and defect channels, with undetermined coefficients. We then impose the superconformal Ward identities (\ref{scfWardid}) to constrain these coefficients. However, several important differences arise in the implementation, which we outline below. Further details will be presented in the companion paper \cite{longpaper}.

\begin{figure}[h]
    \centering
    \begin{subfigure}[b]{0.12\textwidth}
        \centering
        \includegraphics[width=\textwidth]{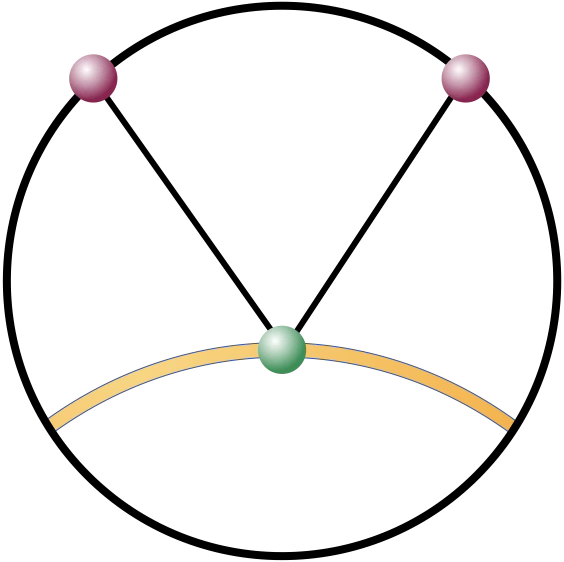}
        \caption{contact}
        \label{fig:WD_con}
    \end{subfigure}
     \hspace{0.02\textwidth}
    \begin{subfigure}[b]{0.12\textwidth}
        \centering
        \includegraphics[width=\textwidth]{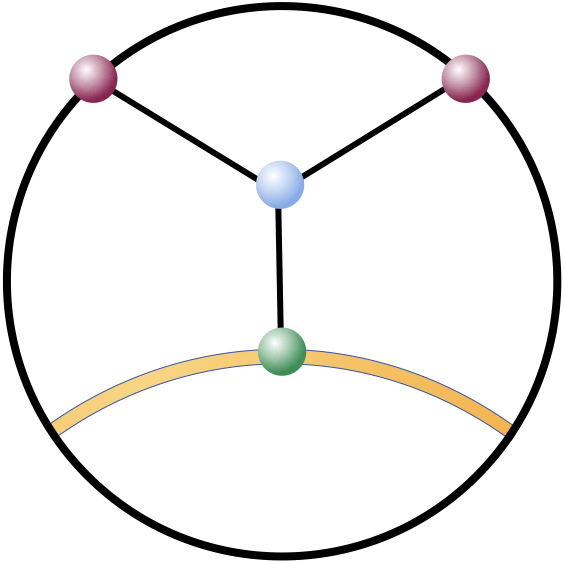}
        \caption{bulk}
        \label{fig:WD_bulk}
    \end{subfigure}
     \hspace{0.02\textwidth}
    \begin{subfigure}[b]{0.12\textwidth}
        \centering
        \includegraphics[width=\textwidth]{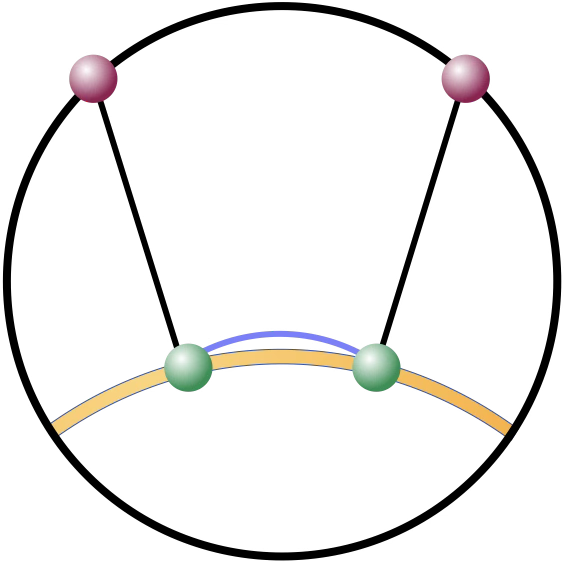}
        \caption{defect}
        \label{fig:WD_defect}
    \end{subfigure}
    \caption{Three types of Witten diagrams at tree-level. The yellow curve represents the giant graviton defect.}
    \label{fig:WD}
\end{figure}

\vspace{0.2cm}
\noindent{\bf Orbital averaging.} Unlike maximal giants, non-maximal and dual giants have an additional orbital modulus $\phi$ parameterizing distinct classical embeddings (Fig. \ref{fig:orbitalplane}). Since the  giant graviton operator has a definite $U(1)$ charge, the averaging over $\phi$ must be performed quantum mechanically, rather than by averaging Witten diagrams at fixed $\phi$ \cite{Chen:2026ium}. It is important to regard $\phi$ as a collective coordinate, not as an additional worldvolume direction \footnote{For earlier studies of orbital averaging in giant graviton correlators, see \cite{Yang:2021kot,Holguin:2022zii,Holguin:2025dei}}.

Expanding in Fourier modes of the orbital angle $\phi$ introduces a tower of $U(1)$ angular momenta labeled by $q\in\mathbb{Z}$. These combine with the intrinsic charge $r$ of defect fluctuations as $r\to r+q$. On the other hand, the motion of the brane induces a Doppler shift between the center-of-mass (CM) and static frames \cite{Chen:2026ium}
\begin{equation}
    \Delta_{\rm static}=\Delta_{\rm CM}+r\;.
\end{equation}
The addition of orbital angular momentum therefore leads to $\Delta_{\rm static}\to\Delta_{\rm static}+q$, enlarging the effective defect spectrum entering the bootstrap ansatz. The allowed values of $q$ are constrained by the branching of the external $SO(6)$ representations under $SO(4)\times U(1)$.

\begin{figure}[h]
    \centering
    \begin{subfigure}[b]{0.15\textwidth}
        \centering
        \includegraphics[width=\textwidth]{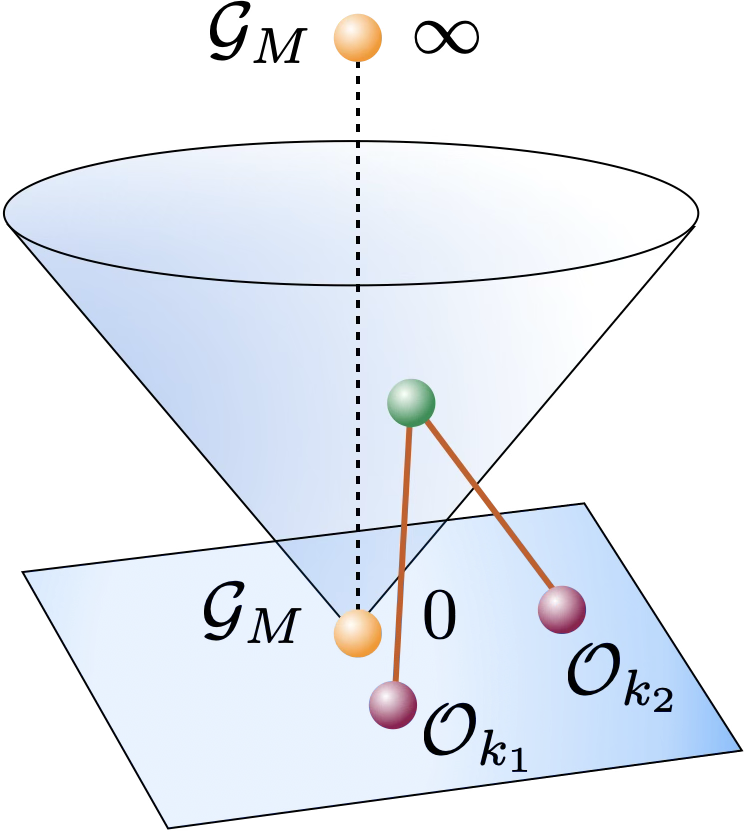}
        \caption{contact}
        \label{fig:dualWD_con}
    \end{subfigure}
     \hspace{0.0\textwidth}
    \begin{subfigure}[b]{0.15\textwidth}
        \centering
        \includegraphics[width=\textwidth]{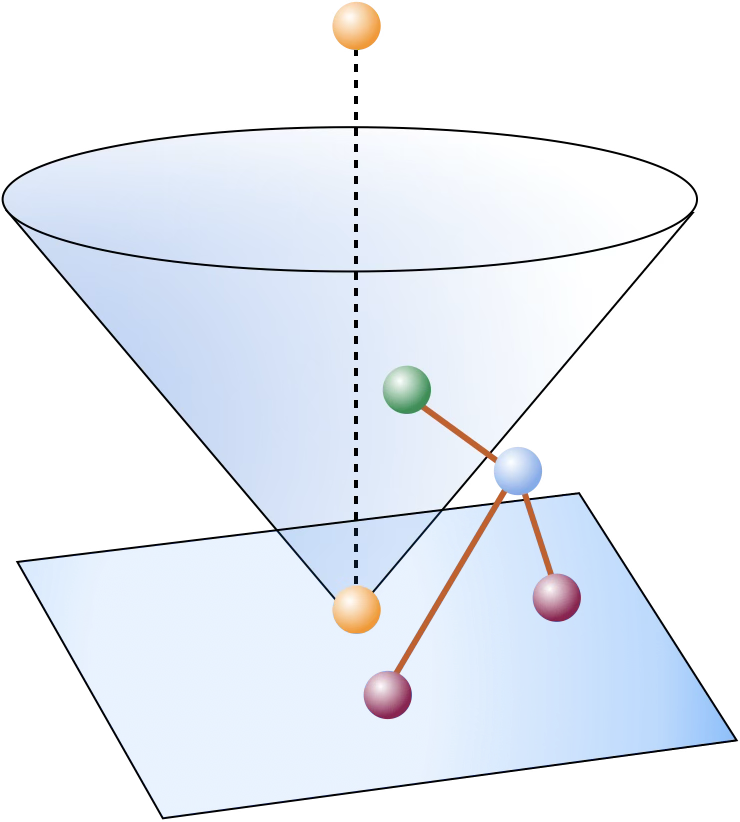}
        \caption{bulk}
        \label{fig:dualWD_bulk}
    \end{subfigure}
     \hspace{0.0\textwidth}
    \begin{subfigure}[b]{0.15\textwidth}
        \centering
        \includegraphics[width=\textwidth]{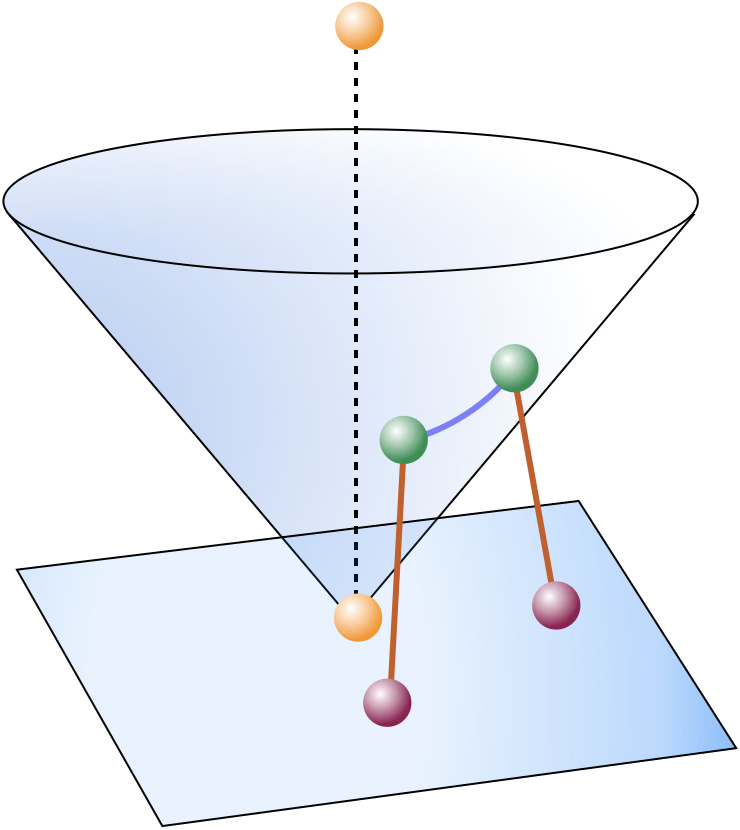}
        \caption{defect}
        \label{fig:dualWD_defect}
    \end{subfigure}
    \caption{Dual giant graviton Witten diagrams in Poincar\'e coordinates. Green vertices are integrated over the cone.}
    \label{fig:dualWD}
\end{figure}

\vspace{0.2cm}
\noindent{\bf New diagrams.} Another complication is diagrammatics. For sphere giants, the wrapped S$^3$ lies inside S$^5$. After Kaluza-Klein reduction, the brane is localized on a geodesic in AdS$_5$, and the Witten diagrams are the same as those in the maximal case \cite{Chen:2025yxg,Chen:2026ium}. For dual giants, however, the S$^3$ is inside AdS$_5$, giving a 4d defect worldvolume that ends at the heavy operator insertions on the boundary. Inserting the operators at $0$ and $\infty$, the defect is a cone in Poincar\'e coordinates, as is shown in Fig. \ref{fig:dualWD}. This leads to a new class of defect Witten diagrams where defect vertices are integrated over a 4d cone geometry rather than along a 1d geodesic. Nevertheless, these more complicated diagrams can be evaluated by extending the techniques of \cite{Rastelli:2017ecj}. For example, the contact Witten diagram with external dimensions $2$ and no derivatives in the vertex evaluates to
\begin{equation}
\nonumber    C_{22}^{\rm D}=\!\frac{2\pi^2\beta^2 V}{\Delta_\beta}\!
\left[
\frac{1\!+\!V\!+\!\beta U}{\sqrt{\Delta_\beta}}
\operatorname{arcosh}\!\left(
\frac{1\!+\!V\!+\!\beta U}{2\sqrt V}
\right)\!-1
\right]\,,
\end{equation}
where $\Delta_\beta=(1+V+\beta U)^2-4V$.
The bulk channel exchange Witten diagrams become finite sums of contact Witten diagrams after using the integrated vertex identities. The defect channel exchange Witten diagrams can be evaluated by using the equation of motion identities. We give explicit examples in the Supplemental Material while postponing a systematic discussion to \cite{longpaper}. 

\vspace{0.2cm}
\noindent{\bf Multiple solutions.} For maximal giants, the superconformal Ward identities fix the ansatz up to an overall constant \cite{Chen:2025yxg,Chen:2026ium}. For non-maximal and dual giants, however, the enlarged defect spectrum leads to
additional solutions, particularly when  $\min\{k_1,k_2\}>2$. We find two types of ambiguities. First, each bulk supergravity multiplet can be completed into an independent solution of (\ref{scfWardid}) by adding defect exchanges and contact diagrams. Second, certain combinations of defect exchanges give rise to purely rational, free-field-like contributions.

\vspace{0.2cm}
\noindent{\bf Bootstrap results.} The remaining ambiguities can be fixed by comparing with (\ref{solscfWardid}) and requiring the free-theory correlator to be reproduced. The bulk exchange coefficients in the bootstrap results can be independently determined from the one-point functions of the exchanged supergravity fields, providing nontrivial consistency checks. We present the results for the simplest $k_1=k_2=2$ cases below, while the higher-weight results will be organized into generating functions in the next section. For non-maximal sphere giants, the reduced correlator is particularly simple
\begin{equation}\label{H22nonmaximal}
    H_{22}^{\rm S}=\eta H_{22}^{\rm max}\;,
\end{equation}
where $H_{22}^{\rm max}$ is the maximal giant reduced correlator \cite{Chen:2025yxg,Chen:2026ium}
\begin{equation}
    H_{22}^{\rm max}=\frac{2V(V^2-2V\log V-1)}{N(P_1\cdot\mathbb N\cdot P_1)^2
(P_2\cdot\mathbb N\cdot P_2)^2U(1-V)^3}\;,
\end{equation}
and we note it can be expressed using only $P_1$, $P_2$ and $\mathbb{N}$. For dual giants, despite the more involved Witten diagrams, the final result takes a compact form
\begin{equation}\label{H22dual}
\begin{aligned}
H_{22}^{\rm D}
{}&\!=\!\frac{1}{(P_1\cdot\mathbb N\cdot P_1)^2
(P_2\cdot\mathbb N\cdot P_2)^2}
\frac{8\beta V^2}{NU \Delta_\beta^{3/2}}
\\
&\!\times\!\!\left[
{\rm arcosh}\left(
\frac{1\!+\!V\!+\!\beta U}{2\sqrt V}
\right)
\!-\!\frac{\sqrt{\Delta_\beta}(1\!+\!V\!+\!\beta U)}{4V}
\right].
\end{aligned}
\end{equation}
As a further consistency check, integrating these reduced correlators with the normalization of \cite{Chen:2025yxg} yields $\eta$ and $\log(1+\beta)$ respectively, in agreement with localization predictions \cite{Brown:2024tru,Brown:2025huy,Brown:2026dhy}.

\section{Hidden conformal symmetry}
Both non-maximal and dual giant graviton correlators turn out to enjoy a higher dimensional symmetry, allowing all correlators to be packaged into simple generating functions via the higher-dimensional uplift of the lowest-weight correlator. These structures are similar to the maximal giant version \cite{Chen:2025yxg,Chen:2026ium}, but with the key distinguishing feature of averaging over the angular parameter $\phi$. In what follows, we will see how the uplifting naturally follows from  geometry, and how both cases can be obtained from the maximal giant one via $SO(2,10)$ rotations.

\vspace{0.2cm}
\noindent{\bf Non-maximal giants.} We first consider the non-maximal giant gravitons. We define a projector in $\mathbb{R}^{2,10}$ that projects to the subspace $\mathcal{S}^{\rm S}_{\eta,\phi}$
\begin{equation}
\Pi_{\rm S}=-\eta^{-1}(E_0^{\rm S}\otimes E_0^{\rm S}+E_1^{\rm S}\otimes E_1^{\rm S})+\sum_{i=1}^4 F_i\otimes F_i\;.
\end{equation}
The R-symmetry projector (\ref{defM}) should be identified here with $\mathbb{M}=\sum_{i=1}^4 F_i\otimes F_i$, or equivalently
\begin{equation}
\mathbb{M}={\bf 1}_R-B_0\otimes B_0-B_1\otimes B_1\;.
\end{equation}
We can relate $B_0$, $B_1$ to $t_3$, $t_4$ via  
\begin{equation}\label{ttoB}
\frac{t_3}{\sqrt{t_3\cdot t_4}}=\frac{B_0-iB_1}{\sqrt{2}}\;,\quad \frac{t_4}{\sqrt{t_3\cdot t_4}}=\frac{B_0+iB_1}{\sqrt{2}}\;,
\end{equation}
which reproduces (\ref{defM}). Similarly, $P_3$, $P_4$ can be expressed in terms of $A_0$, $A_1$ as
\begin{equation}\label{PtoA}
    \frac{P_3}{\sqrt{-P_3\cdot P_4}}=\frac{A_0-iA_1}{\sqrt{2}}\,,\; \frac{P_4}{\sqrt{-P_3\cdot P_4}}=\frac{A_0+iA_1}{\sqrt{2}}\,.
\end{equation}
Let us now define for the light operators
\begin{equation}
\mathcal{Z}_a=\left(P_a,\frac{\lambda_a}{\sqrt{2}}t_a\right)\;,\quad a=1,2\;.
\end{equation}
The uplifting rules for non-maximal giants are
\begin{equation}
   T_{\rm S}= \left\{\begin{array}{c} P_1\cdot P_2 \to \mathcal{Z}_1\cdot \mathcal{Z}_2\\ P_a\cdot \mathbb{N}\cdot P_b \to \eta\, \mathcal{Z}_a\cdot \Pi_{\rm S} \cdot \mathcal{Z}_b
   \end{array}\right\}\;.
\end{equation}
The factor of $\eta$ follows from consistency because restricting to only the AdS components we have
\begin{equation}
    (P_a,0)\cdot \Pi_{\rm S}\cdot (P_b,0)=\frac{1}{\eta} P_a\cdot \mathbb{N}\cdot P_b\;.
\end{equation} 
We find that the reduced correlators of the non-maximal giant gravitons can be packaged into the following generating function
\begin{equation}\label{bfHS}
    {\bf H}_{\rm S}(\eta)=\int_0^{2\pi} \frac{d\phi}{2\pi} T_{\rm S}[H_{22}^{\rm S}]\;.
\end{equation}
Taylor expanding the generating function, we recover
\begin{equation}
    H_{k_1k_2}^{\rm S}=\frac{\sqrt{k_1 k_2}}{2}\mathbf{H}_{\rm S}\big|_{\lambda_1^{k_1-2}\lambda_2^{k_2-2}}\;.
\end{equation}

\vspace{0.2cm}
\noindent{\bf Dual giants.}
For dual giant gravitons, we can similarly define the projector to $\mathcal{S}_{\beta,\phi}^{\rm D}$
\begin{equation}
\Pi_{\rm D}=-\beta^{-1}(E_0^{\rm D}\otimes E_0^{\rm D}+E_1^{\rm D}\otimes E_1^{\rm D})+\sum_{i=1}^4 G_i\otimes G_i\;.
\end{equation}
Here we note $\mathbb{N}=-A_0\otimes A_0-A_1\otimes A_1$ and therefore
\begin{equation}
\sum_{i=1}^4 G_i\otimes G_i={\bf 1}_{\rm AdS}-\mathbb{N}\;.
\end{equation}
The vectors $A_0$, $A_1$ and $B_0$, $B_1$ are related to $P_3$, $P_4$ and $t_3$, $t_4$ in the same way as in the non-maximal giant case, see (\ref{ttoB}) and (\ref{PtoA}). 

We define the uplifting rules for dual giants to be 
\begin{equation}
   T_{\rm D}= \left\{\begin{array}{c} P_1\cdot P_2 \to \mathcal{Z}_1\cdot \mathcal{Z}_2\\ P_a\cdot \mathbb{N}\cdot P_b \to \beta\, \mathcal{Z}_a\cdot (\Pi_{\rm D}-{\bf 1}_{12{\rm d}}) \cdot \mathcal{Z}_b
   \end{array}\right\}\;,
\end{equation}
where ${\bf 1}_{12{\rm d}}$ is the identity in the total embedding space. The reason for having $\beta(\Pi_{\rm D}-{\bf 1}_{12{\rm d}})$ as the analogue of $\eta\Pi_{\rm S}$ is that $\Pi_{\rm D}$ also contains the AdS identity component. More precisely, restricting to the AdS part we have
\begin{equation}
    (P_a,0)\cdot \Pi_D \cdot (P_b,0)=P_a\cdot \left({\bf 1}_{\rm AdS}+\frac{1}{\beta}\mathbb{N}\right) \cdot P_b\;.
\end{equation}
Quite remarkably, the dual giant graviton reduced correlators can also be packaged into a generating function
\begin{equation}\label{bfHD}
    {\bf H}_{\rm D}(\beta)=\int_0^{2\pi} \frac{d\phi}{2\pi} T_{\rm D}[H_{22}^{\rm D}]\;,
\end{equation}
with
\begin{equation}
    H_{k_1k_2}^{\rm D}=\frac{\sqrt{k_1 k_2}}{2} \mathbf{H}_{\rm D}\big|_{\lambda_1^{k_1-2}\lambda_2^{k_2-2}}\;.
\end{equation}

\vspace{0.2cm}
\noindent{\bf Unified via $SO(2,10)$.} The similarity between the two cases can be made more precise. In fact, both can be obtained from the maximal case by $SO(2,10)$ rotations in the total embedding space. Let us first focus on the geometry. We combine the $SO(1,1)$ boost and $SO(2)$ rotation into $B_{\rho,\phi}$ which acts on the vectors as
\begin{equation}
C_i(\rho,\phi)=\cosh\rho A_i+\sinh\rho B_i(\phi)\;,
\end{equation}
with $C_i\cdot C_j=-\delta_{ij}$.
We also define $R_\theta\in SO(8)$ 
\begin{equation}
\nonumber    (F_i(\theta),G_i(\theta))=(\cos\theta G_i+\sin\theta F_i,-\sin\theta G_i+\cos\theta F_i)\;.
\end{equation}
We can then define a 6d subspace
\begin{equation}
\mathcal{S}_{\rho,\phi,\theta}={\rm span}\{C_0,C_1,F_1(\theta),F_2(\theta),F_3(\theta),F_4(\theta)\}\;,
\end{equation}
with the projector
\begin{equation}
\Pi(\rho,\phi,\theta)=-\sum_{i=0}^1 C_i\otimes C_i+\sum_{j=1}^4 F_j(\theta)\otimes F_j(\theta)\;.
\end{equation}
This allows us to start from the maximal giant configuration and reach both non-maximal and dual giant gravitons
\begin{equation}
\mathcal{S}_{\rho,\phi,\theta}=B_{\rho,\phi}\, R_{\theta-\frac{\pi}{2}}\,\mathcal{S}_{0,\phi,\frac{\pi}{2}}\;,
\end{equation}
with the parameters
\begin{equation}
    \begin{split}
        {}&\text{Non-maximal:}\quad \theta=\frac{\pi}{2}\;,\quad {\rm sech}^2\rho=\eta\;,\qquad\\
        {}& \text{Dual:}\quad\quad\quad\quad\;\;\; \theta=0\;,\quad {\rm csch}^2\rho=\beta\;.\qquad
    \end{split}
\end{equation}
To connect with the previous discussion, we note \footnote{We also note that the two boost parameters are related by $\eta=\beta/(1+\beta)$.}
\begin{equation}
E_i^{\rm S}=\sqrt{\eta}\,C_i(\rho,\phi)\;,\quad E_i^{\rm D}=\sqrt{\beta}\,C_i(\rho,\phi)\;.
\end{equation}
 Remarkably, the seed correlators (\ref{H22nonmaximal}) and (\ref{H22dual}) are also related and can be obtained from the maximal giant graviton reduced correlator via
\begin{equation}
    H_{22}^{\rm S/D}=\kappa H_{22}^{\rm max}(P_1,P_2,\mathbb{N})\big|_{\mathbb{N}\to \kappa \Pi_{\rm S/D}}\;,
\end{equation}
where $\kappa$ equals to $\eta$ and $\beta$ for spherical and dual giants respectively. This together with (\ref{bfHS}) and (\ref{bfHD}) give us the unifying formula
\begin{equation}
    {\bf H}_{\rm S/D}(\kappa)=\int_0^{2\pi}\frac{d\phi}{2\pi} \kappa H_{22}^{\rm max}(\mathcal{Z}_1,\mathcal{Z}_2, \kappa \Pi_{\rm S/D}(\phi))\;.
\end{equation}

\section{Discussion and outlook}

Our results reveal a common geometric origin for the correlators of maximal, non-maximal, and dual giant gravitons. A natural question is whether the hidden $SO(2,10)$ relation persists at weak or even finite coupling \footnote{For recent studies of related hidden symmetries in giant graviton correlators at weak coupling, see \cite{Wu:2025ott,He:2026ios}}. This possibility is particularly interesting from the integrability perspective. Maximal giant gravitons correspond to integrable boundary states \cite{Jiang:2019xdz,Jiang:2019zig}, whereas previous studies suggest that integrability is lost for non-maximal and dual giants \cite{Chen:2019gsb}. If the hidden relation continues to hold away from strong coupling, one could exploit integrability in the maximal case to obtain results for non-maximal and dual giants, which would otherwise be much harder to compute.

Another interesting direction is to study corrections beyond tree-level supergravity. Recently, the leading curvature corrections to AdS stringy form factors for maximal giant gravitons were computed using the worldsheet bootstrap \cite{Alday:2026frs}. Extending this analysis to non-maximal and dual giants would allow us to examine whether the hidden geometric structure survives stringy corrections. Independently, one could apply the defect unitarity method of \cite{Chen:2024orp,Chen:2026fnf} to compute $1/N$ corrections to the present results, providing access to quantum gravity corrections to the conformal data of a broader class of defect operators.

It would also be interesting to consider more general correlators, such as four-point functions of giant gravitons \cite{Vescovi:2021fjf,fourgiants} or higher-point functions involving two giants and additional light single-particle operators. Another extension concerns more general heavy $\frac{1}{2}$-BPS operators. Giant gravitons and dual giant gravitons are special cases of Schur polynomial operators \cite{Corley:2001zk,deMelloKoch:2004crq,Aprile:2020uxk}. In particular, one may consider operators whose conformal dimensions scale as $N^2$. From the field theory perspective, the superconformal symmetry remains the same \cite{Kazakov:2024ald}. In the bulk, however, such heavy operators produce backreaction \cite{Abajian:2023jye}, and the resulting geometries can differ substantially from AdS$_5\times$S$^5$, as in the Lin--Lunin--Maldacena backgrounds \cite{Lin:2004nb}. It would be interesting to understand whether similar hidden geometric structures exist in these more general settings.

\vspace{0.3cm}
\noindent{\bf Acknowledgments.} We thank Chang Liu for useful discussions and collaboration on related projects. The work of Q.C., Z.H. and X.Z. is supported by NSFC Grant No. 12275273 and funds from the Chinese Academy of Sciences. Z.H. is also supported by the NSFC Grant No. 12547189. The work of Y.J. is supported by the NSFC through Grant No. 12575073.

\bibliography{refs}
\clearpage
\onecolumngrid
\begin{center}
\large\bfseries Supplemental Material
\end{center}
\setcounter{section}{0}
\setcounter{subsection}{0}
\setcounter{equation}{0}
\setcounter{figure}{0}
\setcounter{table}{0}
\renewcommand{\thesection}{S\arabic{section}}
\renewcommand{\thesubsection}{\thesection.\arabic{subsection}}
\renewcommand{\theequation}{S\arabic{equation}}
\renewcommand{\thefigure}{S\arabic{figure}}
\renewcommand{\thetable}{S\arabic{table}}
\renewcommand{\theHsection}{S\arabic{section}}
\renewcommand{\theHsubsection}{\theHsection.\arabic{subsection}}
\renewcommand{\theHequation}{S\arabic{equation}}
\renewcommand{\theHfigure}{S\arabic{figure}}
\renewcommand{\theHtable}{S\arabic{table}}
\setcounter{secnumdepth}{2}
\allowdisplaybreaks[2]

As representative examples, we record here a few Witten diagrams entering the bootstrap of the dual giant graviton correlators. A systematic study of these diagrams will be presented in \cite{longpaper}.

We use $C_{\Delta_1\Delta_2}^{\rm D}$ to denote the basic contact Witten diagram with two bulk-to-boundary propagators of dimensions $\Delta_1$ and $\Delta_2$ and no derivatives in the contact vertex
\begin{equation}
   C_{\Delta_1\Delta_2}^{\rm D}=(-P_1\cdot \mathbb{N}\cdot P_1)^{\frac{\Delta_1}{2}}(-P_2\cdot \mathbb{N}\cdot P_2)^{\frac{\Delta_2}{2}} \int_{\Sigma_\beta}d^4\xi\,\sqrt{\gamma}\,
K_{\Delta_1}(X(\xi),P_1)
K_{\Delta_2}(X(\xi),P_2)\;.
\end{equation}
Here $\Sigma_\beta$ is the dual giant worldvolume, $\gamma$ is its induced metric, and $K_\Delta$ denotes the bulk-to-boundary propagator
\begin{equation}
   K_\Delta(X,P)=\frac{1}{(-2X\cdot P)^\Delta}\;. 
\end{equation}
All contact integrals required for our bootstrap have the same analytic structure as $C_{22}^{\rm D}$ shown in the main text. For example, we have
\begin{equation}
\begin{split}
    C_{13}^{\rm D}={}&\frac{\pi^2\beta^2\sqrt V}{2\Delta_\beta}\left[1+V+\beta U-\frac{4V}{\sqrt{\Delta_\beta}}\text{arcosh}\left(\frac{1+V+\beta U}{2\sqrt V}\right)\right]\;,\\
    C_{24}^{\rm D}={}&\frac{\pi^2\beta^2 V}{6\Delta_\beta^3}\Bigg[(2+3\beta)\Delta_\beta(\Delta_\beta+12V)-2\beta(1+\beta)U(1+V+\beta U)(\Delta_\beta+30V)\\
    &\quad+\frac{12V}{\sqrt{\Delta_\beta}}\Big[4\beta(1+\beta)U(\Delta_\beta+5V)-(2+3\beta)(1+V+\beta U)\Delta_\beta\Big]\text{arcosh}\left(\frac{1+V+\beta U}{2\sqrt V}\right)\Bigg].
\end{split}
\end{equation}

Using the integrated vertex identities \cite{DHoker:1999mqo,Goncalves:2019znr}, the bulk exchange diagrams can be reduced to finite sums of contact integrals. Taking into account the D3-brane one-point coupling, the exchange of the scalar field dual to $\mathcal{O}_2$ at $k_1=k_2=2$ is given by
\begin{equation}
E_{22}^{s_2,\mathrm D}=-\frac{2}{3\beta U}\left[(1+V+\beta U)C_{22}^{\mathrm{D}}+4\sqrt VC_{13}^{\mathrm{D}}\right]\;.
\end{equation}
Other bulk exchanges can be evaluated similarly. For example, the exchange of the massless vector field in the same $k=2$ multiplet gives
\begin{equation}
E_{22}^{A_2,\mathrm D}=\frac{i(V-1)}{6\beta U}\,C_{22}^{\mathrm D}\;.
\end{equation}

For defect exchange diagrams, we consider the four fluctuations $y^a$, $a=1,\ldots,4$, along the directions of S$^5$ transverse to the brane's orbit. Instead of performing the KK decomposition on the wrapped S$^3$  in AdS$_5$, we work directly with the Green function on the dual giant graviton worldvolume $\mathbb R\times {\rm S}^3$, thereby incorporating the entire tower of modes. The propagator is given by
\begin{equation}
\langle y^a(t_1,\Omega_1) y^b(t_2,\Omega_2)\rangle=\frac{\delta^{ab}}{8\pi^2\left[\cosh(t_1-t_2)-\Omega_1\cdot\Omega_2\right]}\;,
\end{equation}
where $\Omega_i\in {\rm S}^3$. Using this propagator, the transverse exchange diagram at $k_1=k_2=2$ in the $U(1)$ charge sector $m=-1$ evaluates to
\begin{equation}
\begin{split}
\widehat{E}_{22}^{y^a,-1}={}&\frac{32}{3\beta}\,C_{24}^{\mathrm D}+\frac{32(1+\beta)}{9\beta\Delta_\beta}\Bigg[\Big[\sqrt V(1+V+\beta U)+\frac{\Delta_\beta-2V}{\sqrt{V}}\Big]C_{13}^{\mathrm D}+\Big[(1+V+\beta U)-2V\Big]C_{22}^{\mathrm D}\Bigg]\;.
\end{split}
\end{equation}
The remaining defect exchanges can be treated analogously although their explicit expressions are more involved.

\end{document}